\documentclass{article}

\usepackage{soul}
\usepackage{graphicx}
\usepackage[utf8]{inputenc}
\usepackage{hyperref}
\hypersetup{
    colorlinks=true,
    linkcolor = blue,
    urlcolor  = blue,
    citecolor = blue
}

\newcommand{\keyw}[1]{\ul{#1}}

\title{{\Large Workshop Report}\vspace{0.3cm}\\Detection and Classification in Marine Bioacoustics with Deep Learning}
\author{Fabio Frazao, Bruno Padovese, Oliver S.\ Kirsebom }
\date{December 2019}

\begin{document}

\maketitle

On 21--22 November 2019, about 30 researchers gathered in Victoria, BC, for the workshop {\it Detection and Classification in Marine Bioacoustics with Deep Learning} organized by MERIDIAN and hosted by Ocean Networks Canada. The workshop was attended by marine biologists, data scientists, and computer scientists coming from both Canadian coasts and the US and representing a wide spectrum of research organizations including universities, government (Fisheries and Oceans Canada, National Oceanic and Atmospheric Administration), industry (JASCO Applied Sciences, Google, Axiom Data Science), and non-for-profits (Orcasound, OrcaLab). Consisting of a mix of oral presentations, open discussion sessions, and hands-on tutorials, the workshop program offered a rare opportunity for specialists from distinctly different domains to engage in conversation about Deep Learning and its promising potential for the development of detection and classification algorithms in underwater acoustics. Presentations given at the workshop can be found on MERIDIAN's website at \href{https://meridian.cs.dal.ca/}{meridian.cs.dal.ca}, while the hands-on tutorial can be found at \href{https://gitlab.meridian.cs.dal.ca/workshops/victoria\_nov2019}{gitlab.meridian.cs.dal.ca/workshops/victoria\_nov2019}. In the following, we summarize key points from the presentations and discussion sessions. The list of participants can be found at the end of the report. 

\newpage

\begin{figure}[h]
\centering
\includegraphics[width=\textwidth]{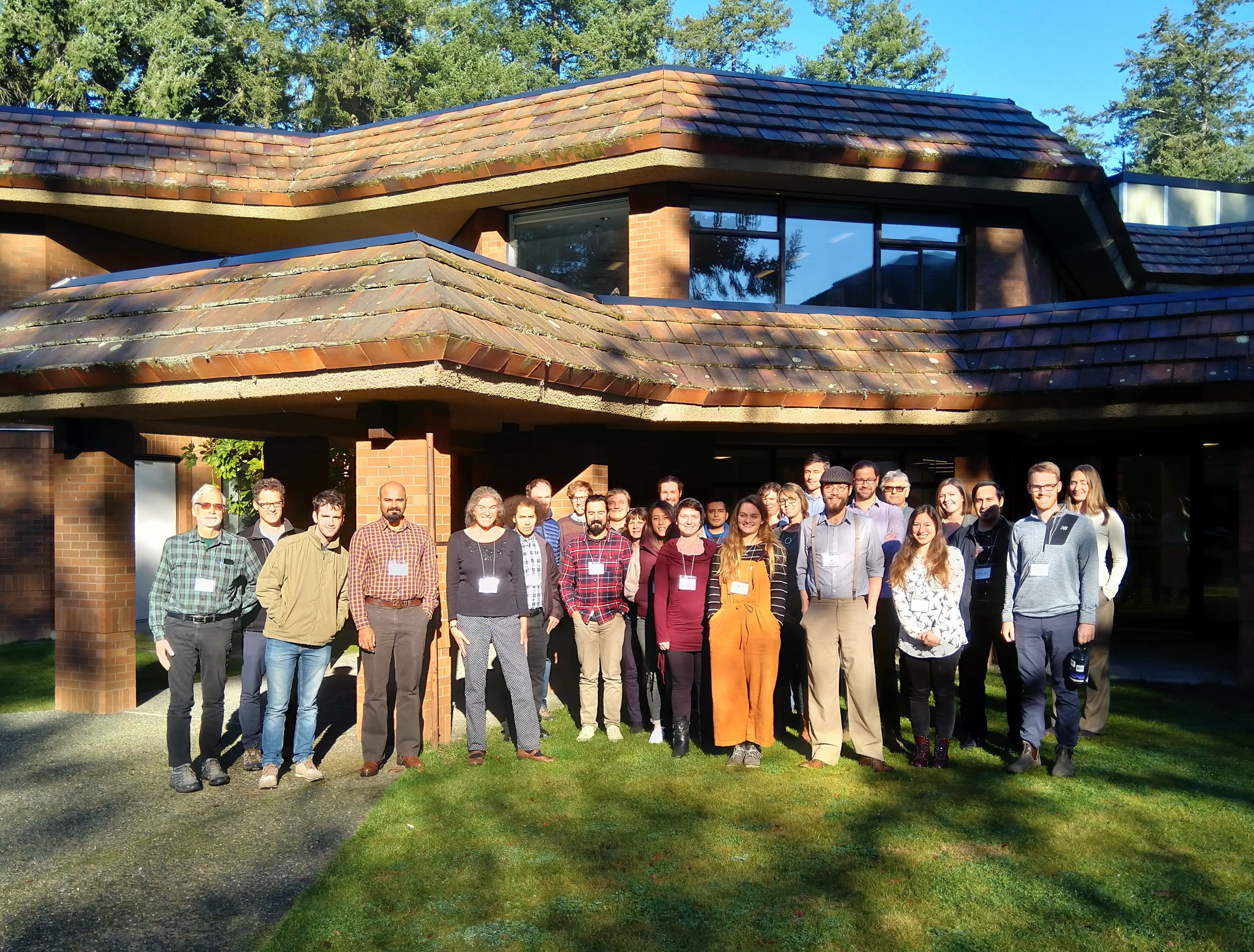}
\caption{Attendees at the workshop {\it Detection and Classification in Marine Bioacoustics with Deep Learning} held on 21--22 November 2019 at Ocean Networks Canada in Victoria BC, Canada.}
\end{figure}

\newpage

\section{Deep Learning in a Nutshell}

Introductions to Machine Learning and Deep Learning were given on the first day of the workshop by Marie Roch (San Diego State University) and Fabio Frazao (MERIDIAN). %

\keyw{Machine Learning} (ML) is an approach to creating Artificial Intelligence systems. Its defining characteristic is the use of algorithms capable of learning to perform a task from data, as opposed to having every single step explicitly programmed. These algorithms may act either directly on the raw data or on a (heavely) reduced \keyw{representation} of the data. For many years, the latter was the favored (and only feasible) alternative, but this has changed within the last decade. Aided by growth in computational power and increased availability of training data, algorithms have been developed that successfully learn patterns and correlations directly from raw data, giving birth to a new sub-discipline of ML known as \keyw{Deep Learning} (DL)~\cite[Chapter~1]{goodfellow2016}. %

Thus, the defining characteristic of a DL model is that it acts on the raw data, or a high-dimensional representation very close to the raw data, rather than a small set of \emph{features}. By employing a layered structure, DL models are capable of learning the representation that is best suited for solving a given task (adaptability), and the same DL model may be trained to solve multiple tasks (versatility). Moreover, knowledge acquired in one domain may be transfered to solve a related task in a different domain (transferability). %
Compared to convetional ML models, DL models tend to be data-hungry and computationally intensive to train (but not necessarily to run). 

All DL models employ some type of \keyw{Artificial Neural Network}~\cite[Chapter~6]{goodfellow2016}. Within the last decade, neural networks have become the preferred ML algorithm for solving a wide range of tasks, outperforming existing methods and achieving human-level accuracy in domains such as image analysis~\cite{he2015} and natural speech processing \cite{hinton2012}. Originally inspired by the human brain, neural networks consist of a large number of interconnected ``neurons'', each typically performing a simple linear operation on input data, specified by a set of weights and a bias, followed by an activation function. In a \keyw{supervised training} approach, the network is given examples of labeled data, and the weights and biases are optimized using a gradient-descent technique. 

Modern neural networks exhibit multi-layer architectures, which enable them to build complex concepts out of simpler concepts and hence learn a non-linear representation of the data. Therefore, modern neural networks are often referred to as \keyw{Deep Neural Networks}. Two of the most commonly encountered basic architectures are \keyw{Convolutional} Neural Networks (CNNs) and \keyw{Recurrent} Neural Networks (RNNs), which are particularly well adapted to analyzing image data and sequential data, respectively~\cite[Chapter~9 \& 10]{goodfellow2016}. %
The availability of large labeled datasets, containing millions of labeled examples, has been a key factor in the success of deep neural networks in domains such as image analysis and natural speech processing. Therefore, much of the current research focuses on how to train deep neural networks more efficiently on smaller datasets~\cite{towardsdatascience-few-shot}. %

\section{Applications to Marine Bioacoustics}

Shallow neural networks have been employed for the purpose of sound detection and classification in marine bioacoustics since the 1990s, usually combined with a method of feature extraction~\cite{bahoura2010}, but also acting directly on the spectrogram~\cite{halkias2013}. In the last few years, the first studies employing modern (i.e.\ deep) neural networks have been reported. Applications include classification of fish sounds~\cite{malfante2018}, detection of orca~\cite{bergler2019, google-orca} and humpback~\cite{google-humpback} vocalizations, classification of calls of multiple whale species~\cite{mcquay2017, thomas2019}, detection of sperm whale echolocation clicks and classification of their codas~\cite{bermant2019}, and detection of the upcall of the North Atlantic right whale~\cite{shiu2020, kirsebom2020}. For a recent review of ML applications to acoustics, see Ref.~\cite{bianco2019}.

At the workshop, presentations given by Herv\'e Glotin (Université de Toulon), Shyam Madhusudhana (Cornell University), Mark Thomas (Dalhousie University), Fabio Frazao (MERIDIAN), Ann Allen (NOAA), and Lauren Harrell (Google), gave an overview of the state-of-the-art. %
A common theme was the use of CNNs acting on spectrogram representations of acoustic data. %

The presentations given by Shyam Madhusudhana and Mark Thomas showcased two complementary and equally relevant approches: Reduce the computational load (both during training and inferrence) as much as possible while maintaining high performance (``DenseNet''), or maximize the performance by building more sophisticated models trained on high-performance computers (``Mask R-CNN''). 
Meanwhile, the presentation given by Fabio Frazao highlighted the importance of having a training dataset with large variance. Indeed, not only the size of the training dataset, but also its variance is a key factor in determining the DL model's ability to \keyw{generalize}, i.e., how well the model will perform on unseen test data. %

Where most presentations were concerned with the detection and classification of low-frequency calls of baleen whales from spectrogram representations, e.g., the upcall of the North Atlantic right whale, Herv{\'e} Glotin discussed a DL model for classifying echolocation clicks of ten different species of whales and dolphins from the raw acoustic waveform. This work highlights the \keyw{complementary nature of the time- and frequency domains}, the former being better suited for short-duration, broad-band signals (e.g.\ clicks) and the latter better suited for longer-duration, narrow-band signals (e.g.\ tonals). For intermediate cases, it is conceivable that a hybrid approach, exploiting both domains simultaneously, may yield the best performing model. %
Another interesting aspect of this work was the investigation of the impact of \keyw{signal-to-noise ratio} (SNR) on the classification accuracy: Synthetic noise was superimposed on the original signal in varying amounts to simulate a range of SNRs, and the change in accuracy was determined. Such a characterization, while restricted to flat noise backgrounds (as opposed to transients), is essential for normalizing model results and enabling inferrences on an absolute level concerning the presence and distribution of marine species. %
Finally, the importance of \keyw{sound directionality} was also addressed. High-frequency sounds, in contrast to low-frequency sounds, tend to be highly anisotropic, implying that a receiver positioned in front of the sound-producing animal will record a very different sound from a receiver positioned behind it or below it. On one hand, this makes the classification task more challenging. On the other hand, it implies that directional information is embedded in the waveform and a DL model could potentially learn to infer directionality.

A limitation common to all the DL models presented at the workshop, shared by both time- and frequency-domain approaches, is the neglect of \keyw{long-term temporal context}. For example, the spectrograms used as input for the CNNs had lengths ranging from just a few seconds to at most a few tens of seconds. Going forward, the integration of long-term temporal context (as commonly done by human analysts) could prove critical for further enhacing the performance of DL models. This may require that we give up our cherised CNNs in favour of more sophisticated network architectures. The use of \keyw{Convolutional Recurrent Networks}~\cite{choi2017convolutional} as \keyw{Seq2Seq} models~\cite{bahdanau2014neural, luong2015multi} provides a promising approach to solving this challenge. %

Another limitation, common to all frequency-domain models, is the dependence on the choice of \keyw{transform} (Fourier or wavelet) and spectrogram \keyw{resolution}. The issue here is not necessarily one of reduced performance, but loss of versatility and transferability. For example, spectrogram parameters that produce enhanced signal-to-noise ratio for North Atlantic right whale upcalls may not be optimal for calls of other whale species. Various methods to mitigate this issue have been proposed, including the use of a superposition of several different spectrogram representations~\cite{thomas2019} and models that learn the optimal set of wavelets from the data~\cite{balestriero2018}.

Presentations given by Herv\'e Glotin and Bruno Padovese (MERIDIAN) discussed various \keyw{augmentation} techniques for synthetically enlarging the size of a training dataset to produce better-performing DL models. Such techniques range from the simple to the sophisticated. Simple augmentation techniques include: adding synthetic noise to a signal, producing several copies of the same signal that differ only by time shifts, and removing selected pixels or whole regions of a spectrogram~\cite{Salamon2017}. Sophisticated augmentation techniques include training \keyw{Generative Adversarial Networks} (GANs)~\cite{towardsdatascience-GANs} to create synthetic signals and using sound propagation models to account for acoustic underwater effects such as reverberations and multi-path echoes. We note that GANs have been successfully employed to create highly realistic synthetic music~\cite{huang2018timbretron}, suggesting that similar results could be achieved in underwater acoustics. 

The potential benefits of \keyw{Transfer Learning}~\cite{weiss2016survey} were also discussed briefly at the workshop, but due to a lack of studies on Transfer Learning in the context of underwater acoustics little could be concluded. It remains to be seen if Transfer Learning can help boost performance in cases where the amounts of training data are limited. That said, it seems likely that we will be able to use Transfer Learning to great effect to reduce training time and computational demand.

Finally, \keyw{semi-supervised} and \keyw{unsupervised} approaches were identified as techniques that could help leverage unlabelled data for training purposes (see, e.g., Ref.~\cite{balestriero2017}) and be used for characterizing complex soundscapes.

\section{A Community Effort}

Public datasets such as ImageNet have been instrumental in driving progress in computer vision and related domains. Such datasets both facilitate model development and serve as \keyw{benchmarks} enabling inter-model comparisons. Moreover, public datasets have been extensively used as part part of \keyw{challenges}, further incentivizing model development. 

Several factors contribute to the usability and impact of public datasets such as ImageNet. Most importantly, the \keyw{annotations} must of be of a hiqh quality and the \keyw{documentation} provided must be clear and sufficient. Moreover, the data must be easily \keyw{accessible} and remain accessible for many years to ensure continuity and allow the community to track long-term progress. Finally, there needs to be clearly defined \keyw{tasks and performance metrics}, allowing inter-model comparisons.

In marine bioacoustics, the DCLDE (Detection, Classification, Localization and Density Estimation) workshop series (2003--2020) has been the main producer of public datasets:  
\begin{itemize}
\item Workshops 1--6: \href{http://www.mobysound.org/workshops.html}{http://www.mobysound.org/workshops.html}
\item Workshop 7: \href{http://www.cetus.ucsd.edu/dclde/dataset.html}{http://www.cetus.ucsd.edu/dclde/dataset.html}
\item Workshop 8: \href{http://sabiod.univ-tln.fr/DCLDE/challenge.html}{http://sabiod.univ-tln.fr/DCLDE/challenge.html}
\item Workshop 9: \href{http://www.soest.hawaii.edu/ore/dclde/}{http://www.soest.hawaii.edu/ore/dclde/}
\end{itemize}
These datasets, despite certain shortcomings discussed below, have been and continue to be an important resource for the community, facilitating the development of improved detection and classification algorithms. The fact that DCLDE datasets played an important role in three of the presentations given a this workshop (Herv{\'e} Glotin, Shyam Madhusudhana, Fabio Frazao) testifies to this fact.

While there are known issues with the annotation quality of some of the DCLDE datasets and minor gaps in the documentation (esp.\ concerning annotation protocols), the datasets have other, more significant, shortcomings: The datasets often lack clearly defined tasks and performance metrics and do not always remain available after the workshops (e.g.\ workshop no.\ 8). The lack of a single platform and/or a unified standard format for sharing the datasets further reduces their value.

The DCASE (Detection and Classification of Acoustic Scenes and Events) workshop series (2013--2019) provides an excellent example of a \keyw{community platform} that allows exchange of datasets, model results, and tracking of progress over the years  (\href{http://dcase.community/}{http://dcase.community/}). 
Another platform of interest is Kaggle, which may even be a viable option for sharing an annotated acoustic dataset and hosting challenges.

Hosting of \keyw{large datasets} ($> 1$~TB) is a challenge that needs to be addressed by the community. Collaborations with large tech companies such as Google provide one viable solution, as demonstrated by the presentations given by Ann Allen (NOAA) and Lauren Harrell (Google), and is also the solution adopted for the upcoming DCLDE Workshop no.\ 9, where the dataset has a size of 8~TB. However, while such large datasets may be needed to achieve the best possible performance, it must be remembered that smaller datasets of tens of GB or less often are sufficient for model development and benchmarking. This is particularly true for data acquired at low sampling rates suitable for classification of low-frequency species.

It is difficult to set down specific guidelines for the creation of training and test sets, but some general observations can still be made: Annotations must (of course) have high quality, but more importantly, the \keyw{annotation protocol} must be clearly documented: Were the files annotated at the call level and, if so, were all calls annotated or only one representative call per file? What level of certainty was required from the analyst? Did the analyst inspect/listen to the entire recordings, or did the analyst only validate the results of an automated detection algorithm, etc. %
Datasets should always be accompanied by \keyw{clearly defined tasks}, which should be aligned with the interests of the marine bioacousticians and the quality and characteristics of the annotations available. Even weakly labelled data can be useful for model development, if the tasks are well-aligned with the data. 
Datasets should contain a reasonably balanced proportion of samples from each class of interest, and should ideally cover the full spectrum of \keyw{acoustic conditions} likely to be encountered in applications. That is, the larger the variance of the dataset, the better. However, if the intention is to develop a model that will only be used in restricted geographical region, or a restricted period of the year (say, the summer months), then it is perfectly fine and useful to have datasets that focus on these regions and time periods.

\keyw{Licensing} must also be considered. While it would be desirable to enforce \keyw{open-sourcing} of models derived from public datasets, this could prove counter productive in terms of incentivising model development.

Creating a high-quality training dataset requires a lot of effort. Luckily, we are seeing more \keyw{incentives} to undertake such efforts as journals are opening for submissions of pure datasets, e.g., Nature Scientific Data and Data in Brief. Also, labs and companies can always choose to release only partial datasets, thereby maintaining a competitive advantage by being able to train their own models on a larger dataset than the one available to the public.

\section{From Academic Exercises to Practical Tools}

As discussed above, Deep Neural Networks exhibit a high degree of flexibility, i.e., the same network architecture can be trained to solve a range of different tasks. While it requires a good understanding of Machine Learning to design and implement a neural network, training it is, in principle, straightforward. All you must do is feed the appropriate training data to the neural network. Therefore, it is conceivable that not just experts, but also people with limited programming experience and limited knowledge of machine learning could develop or re-purpose DL-based detection and classification algorithms. However, to achieve this ambitious goal, the proper \keyw{software tools} first need to be developed. %
MERIDIAN is contributing towards this goal through the development of the open-source Python library Ketos (\href{https://docs.meridian.cs.dal.ca/ketos/}{docs.meridian.cs.dal.ca/ketos/}). At the workshop, participants could explore the functionalities of Ketos in a 90-min hands-on tutorial (\href{https://gitlab.meridian.cs.dal.ca/workshops/victoria_nov2019}{gitlab.meridian.cs.dal.ca/workshops/victoria\_nov2019}).

Marine biologists in general, but especially those working in government research, are under constant pressure to provide marine environmental regulators with data, reliable results, and firm conclusions. Since conditions and requirements imposed by regulators often change, it would be useful to have \keyw{adaptable} detection and classification algorithms. However, adaptability is not the only desirable feature. Equally important, is improved \keyw{performance} and the capacity to provide \keyw{uncertainty estimation}. DL models have the potential to meet all of these requirements.

While achieving human-level performance remains the number one long-term goal, producing adaptable models is an important complementary goal, which could be achieved in shorter time. This will require the development of \keyw{graphical user interfaces} (GUI) since few users will be comfortable with programming. With a GUI, users could interact with DL models without having to do any programming, but some understanding and control of machine-learning parameters will still be required for the successful implementation of such an \keyw{Active Learning} scheme~\cite{settles2011theories}.

Some level of \keyw{compatability} with existing software, e.g., Pamguard and RavenPro, is also highly desirable. As a minimum this implies ensuring that input/output formats are compatible, i.e., that a list of detections produced by the DL software can be loaded into commonly used annotation software tools for review (and editing), and conversely that annotation tables produced by the same commonly used annotation tools can be loaded into the DL software. More advanced levels of compatability would involve the development of plug-ins for existing software, such as the Humpback whale detector plug-in developed by Google for use in Pamguard by NOAA, as discussed in Ann Allen's presentation. Another practical concern is to ensure functionality across OS systems (Windows, Mac, Linux) and dealing with strict firewalls present especially in government research labs.

As a final note, we mention the VIAME toolkit (\href{http://www.viametoolkit.org/}{www.viametoolkit.org}) as a recent example of an open-source toolkit allowing users to train ML models (for underwater image processing, not audio) through a GUI without the need to program.

\section{Public Sharing of Data and Software}

Presentations addressing (among other things) public sharing of acoustic data were given by Scott and Val Veirs (Orcasound), Ann Allen (NOAA), and Lauren Harrell (Google). 

Streaming of stored or live acoustic data is a convenient method of \keyw{data sharing}, particularly well suited for outreach activities. If combined with a public \keyw{annotation platform}, live streaming provide a means to build training datasets (assuming the acoustic data is stored somewhere) as currently pursued by Orcasound (\href{http://www.orcasound.net/}{www.orcasound.net}).

Sharing of large datasets, such as NOAA's HARP Data Archive for the Pacific ($> 300$~TB, and growing at 50 TB/yr), is a huge challenge. Ann Allen and Lauren Harrell demonstrated how Google (AI for Social Good) can help solve such problems by providing \keyw{cloud storage, computing power, and analytical tools}. 
It is likely that cloud/decentralized storage and computing will play a growing role in underwater acoustics data handling and analysis in the future. Funding for such services should be built into grant applications. The problem of long-term storage (beyond the duration of the grant), however, remains. 

With scientific journals beginning to accept pure data articles, incentives for sharing data are growing, but this also imposes stricter requirements on data descriptors, which we should welcome as positive development.

For software there is the added challenge of ensuring that the software (and all its dependencies) still runs 5--10 years after it was developed. Adoption by the community of good \keyw{software development practices} should be encouraged, including collaborative coding on platforms like github and gitlab, version control, etc.\ On the other hand, enforcing use of certain tools over others is not realistic, nor productive, so we recommend to go with what works for your problem, but be mindful of what other people in your community.

\section{Conclusion}

We end the report with a brief set of recommendations for the community,
\begin{itemize}
\item Efforts should go into developing models that require less training data,
\item and should be paralleled by efforts towards developing improved data augmentation techniques.
\item Inclusion of long-term temporal context will likely be needed to achieve human-level accuracy. 
\item We also need ``light-weight'' models with minimal power consuption that can be deployed on field instruments.
\item Availability of more and better public datasets is essential to ensure progress.
\item A community platform for hosting datasets, documentation, and task descriptions, and sharing results and code would go a long way!
\item The creation of graphical user interfaces could enable non-programmers to train their own DL models, producing detection algorithms that are tailored to their precise needs.
\end{itemize}


\begin{thebibliography}{10}

\bibitem{goodfellow2016}
Ian Goodfellow, Yoshua Bengio, and Aaron Courville.
\newblock Deep learning book.
\newblock {\em MIT Press}, 521(7553):800, 2016.

\bibitem{he2015}
K.~{He}, X.~{Zhang}, S.~{Ren}, and J.~{Sun}.
\newblock Delving deep into rectifiers: Surpassing human-level performance on
  imagenet classification.
\newblock In {\em 2015 IEEE International Conference on Computer Vision
  (ICCV)}, pages 1026--1034, Dec 2015.

\bibitem{hinton2012}
G.~{Hinton}, L.~{Deng}, D.~{Yu}, G.~E. {Dahl}, A.~{Mohamed}, N.~{Jaitly},
  A.~{Senior}, V.~{Vanhoucke}, P.~{Nguyen}, T.~N. {Sainath}, and
  B.~{Kingsbury}.
\newblock Deep neural networks for acoustic modeling in speech recognition: The
  shared views of four research groups.
\newblock {\em IEEE Signal Processing Magazine}, 29(6):82--97, Nov 2012.

\bibitem{towardsdatascience-few-shot}
\href{https://towardsdatascience.com/advances-in-few-shot-learning-a-guided-tour-36bc10a68b77}{https://towardsdatascience.com/advances-in-few-shot-learning-a-guided-tour-36bc10a68b77}.

\bibitem{bahoura2010}
Mohammed Bahoura and Yvan Simard.
\newblock Blue whale calls classification using short-time fourier and wavelet
  packet transforms and artificial neural network.
\newblock {\em Digital Signal Processing}, 20(4):1256 -- 1263, 2010.

\bibitem{halkias2013}
Xanadu~C. Halkias, S{\'e}bastien Paris, and Herv{\'e} Glotin.
\newblock Classification of mysticete sounds using machine learning techniques.
\newblock {\em The Journal of the Acoustical Society of America},
  134(5):3496--3505, 2013.

\bibitem{malfante2018}
Marielle Malfante, Omar Mohammed, Cedric Gervaise, Mauro Dalla~Mura, and
  Jerome~I. Mars.
\newblock {Use of deep features for the automatic classification of fish
  sounds}.
\newblock In {\em {OCEANS'18 MTS/IEEE}}, Kobe, Japan, May 2018.

\bibitem{bergler2019}
Christian Bergler, Hendrik Schr{\"o}ter, Rachael~Xi Cheng, Volker Barth,
  Michael Weber, Elmar N{\"o}th, Heribert Hofer, and Andreas Maier.
\newblock {ORCA}-{SPOT}: An automatic killer whale sound detection toolkit
  using deep learning.
\newblock {\em Scientific Reports}, 9(1):10997, July 2019.

\bibitem{google-orca}
\href{https://www.blog.google/technology/ai/protecting-orcas/}{https://www.blog.google/technology/ai/protecting-orcas/}.

\bibitem{google-humpback}
\href{https://ai.googleblog.com/2018/10/acoustic-detection-of-humpback-whales.html}{https://ai.googleblog.com/2018/10/acoustic-detection-of-humpback-whales.html}.

\bibitem{mcquay2017}
C.~{McQuay}, F.~{Sattar}, and P.~F. {Driessen}.
\newblock Deep learning for hydrophone big data.
\newblock In {\em 2017 IEEE Pacific Rim Conference on Communications, Computers
  and Signal Processing (PACRIM)}, pages 1--6, Aug 2017.

\bibitem{thomas2019}
Mark Thomas, Bruce Martin, Katie Kowarski, Briand Gaudet, and Stan Matwin.
\newblock Marine mammal species classification using convolutional neural
  networks and a novel acoustic representation.
\newblock {\em arXiv preprint arXiv:1907.13188}, 2019.

\bibitem{bermant2019}
Peter~C. Bermant, Michael~M. Bronstein, Robert~J. Wood, Shane Gero, and
  David~F. Gruber.
\newblock Deep machine learning techniques for the detection and classification
  of sperm whale bioacoustics.
\newblock {\em Scientific Reports}, 9(1):12588, August 2019.

\bibitem{shiu2020}
Y. Shiu, K. J. Palmer, M. A. Roch, {\it et al.}
\newblock Deep neural networks for automated detection of marine mammal species. 
\newblock {\em Scientific Reports}, 10, 607, 2020.

\bibitem{kirsebom2020}
Oliver S.\ Kirsebom, Fabio Frazao, Yvan Simard, Nathalie Roy, Stan Matwin, Samuel Giard.
\newblock Performance of a Deep Neural Network at Detecting North Atlantic Right Whale Upcalls.
\newblock {\em arXiv preprint arXiv:2001.09127}, 2020.

\bibitem{bianco2019}
Michael~J. Bianco, Peter Gerstoft, James Traer, Emma Ozanich, Marie~A. Roch,
  Sharon Gannot, and Charles-Alban Deledalle.
\newblock Machine learning in acoustics: Theory and applications.
\newblock {\em The Journal of the Acoustical Society of America},
  146(5):3590--3628, 2019.

\bibitem{choi2017convolutional}
Keunwoo Choi, Gy{\"o}rgy Fazekas, Mark Sandler, and Kyunghyun Cho.
\newblock Convolutional recurrent neural networks for music classification.
\newblock In {\em 2017 IEEE International Conference on Acoustics, Speech and
  Signal Processing (ICASSP)}, pages 2392--2396. IEEE, 2017.

\bibitem{bahdanau2014neural}
Dzmitry Bahdanau, Kyunghyun Cho, and Yoshua Bengio.
\newblock Neural machine translation by jointly learning to align and
  translate.
\newblock {\em arXiv preprint arXiv:1409.0473}, 2014.

\bibitem{luong2015multi}
Minh-Thang Luong, Quoc~V Le, Ilya Sutskever, Oriol Vinyals, and Lukasz Kaiser.
\newblock Multi-task sequence to sequence learning, 2015.

\bibitem{balestriero2018}
Randall Balestriero, Romain Cosentino, Herve Glotin, and Richard Baraniuk.
\newblock Spline filters for end-to-end deep learning.
\newblock In Jennifer Dy and Andreas Krause, editors, {\em Proceedings of the
  35th International Conference on Machine Learning}, volume~80 of {\em
  Proceedings of Machine Learning Research}, pages 364--373, Stockholmsmässan,
  Stockholm Sweden, 10--15 Jul 2018. PMLR.

\bibitem{Salamon2017}
J.~{Salamon} and J.~P. {Bello}.
\newblock Deep convolutional neural networks and data augmentation for
  environmental sound classification.
\newblock {\em IEEE Signal Processing Letters}, 24(3):279--283, March 2017.

\bibitem{towardsdatascience-GANs}
\href{https://towardsdatascience.com/understanding-generative-adversarial-networks-gans-cd6e4651a29}{https://ai.googleblog.com/2018/10/acoustic-detection-of-humpback-whales.html}.

\bibitem{huang2018timbretron}
Sicong Huang, Qiyang Li, Cem Anil, Xuchan Bao, Sageev Oore, and Roger~B.
  Grosse.
\newblock Timbretron: A wavenet(cycle{GAN}({CQT}(audio))) pipeline for musical
  timbre transfer.
\newblock In {\em International Conference on Learning Representations}, 2019.

\bibitem{weiss2016survey}
Karl Weiss, Taghi~M Khoshgoftaar, and DingDing Wang.
\newblock A survey of transfer learning.
\newblock {\em Journal of Big data}, 3(1):9, 2016.

\bibitem{balestriero2017}
Randall Balestriero, Vincent Roger, Herve~G. Glotin, and Richard~G. Baraniuk.
\newblock Semi-supervised learning via new deep network inversion, 2017.

\bibitem{settles2011theories}
Burr Settles.
\newblock From theories to queries: Active learning in practice.
\newblock In {\em Active Learning and Experimental Design workshop In
  conjunction with AISTATS 2010}, pages 1--18, 2011.

\end{thebibliography}

\newpage

\section{List of Participants}

\begin{table}[h!]
\centering
\begin{tabular}{ |l|l| } 
 \hline
 Name & Affiliation \\
 \hline
 Alex Harris & SFU \\
 Ann Allen & NOAA \\
 Bruno Padovese & MERIDIAN, Dalhousie University \\
 Clair Evers & DFO \\ 
 Dany Alejandro Cabrera Vargas & ONC \\
 Fabio Frazao & MERIDIAN, Dalhousie University \\
 Harald Yurk & DFO \\
 Herv\'e Glotin & Université de Toulon \\
 Hilary Moors-Murphy & DFO \\
 James Pilkington & DFO \\
 Jesse Lopez & Axiom Data Science \\
 Kristen Kanes & ONC \\
 Lauren Harrell & Google \\
 Lisa Leung & MERIDIAN, UBC \\
 Marie Roch & San Diego State University \\
 Mark Thomas & Dalhousie University, JASCO \\
 Miguel Neves Dos Reis & DFO \\
 Oliver Kirsebom & MERIDIAN, Dalhousie University \\
 Quinlan McIntire & OrcaLab \\
 Scott Veirs & Orcasound \\
 Shyam Madhusudhana & Cornell University \\
 Steven Bergner & MERIDIAN, SFU \\
 Suzie Hall & OrcaLab \\
 Val Veirs & Orcasound \\
 Valentina Staneva & University of Washington \\
 Wilfried Beslin & DFO \\
 Xavier Mouy & JASCO \\
 \hline
\end{tabular}
\caption{Attendees at the workshop {\it Detection and Classification in Marine Bioacoustics with Deep Learning} held on 21--22 November 2019 at Ocean Networks Canada in Victoria BC, Canada.}
\label{table:1}
\end{table}

\end{document}